\documentclass{article}

\usepackage{arxiv}

\usepackage[utf8]{inputenc} % allow utf-8 input
\usepackage[T1]{fontenc}    % use 8-bit T1 fonts
\usepackage{hyperref}       % hyperlinks
\usepackage{url}            % simple URL typesetting
\usepackage{booktabs}       % professional-quality tables
\usepackage{amsfonts}       % blackboard math symbols
\usepackage{nicefrac}       % compact symbols for 1/2, etc.
\usepackage{microtype}      % microtypography
\usepackage{lipsum}		% Can be removed after putting your text content
\usepackage{graphicx}
\usepackage{doi}

\usepackage{enumitem}

\title{Computational Inference in Cognitive Science: Operational, Societal and Ethical Considerations} % priority

\author{Baihan Lin\\
	Department of Neuroscience\\
	Columbia University\\
	New York, NY 10027 \\
	\texttt{baihan.lin@columbia.edu} \\
}

\renewcommand{\headeright}{}
\renewcommand{\undertitle}{Technical Report}
\hypersetup{
pdftitle={Computational Inference in Cognitive Science: Operational, Societal and Ethical Considerations},
pdfsubject={q-bio.NC, q-bio.QM},
pdfauthor={Baihan Lin},
pdfkeywords={First keyword, Second keyword, More},
}

\begin{document}
\maketitle

\begin{abstract}

Emerging research frontiers and computational advances have gradually transformed cognitive science into a multidisciplinary and data-driven field. As a result, there is a proliferation of cognitive theories investigated and interpreted from different academic lens and in different levels of abstraction. We formulate this applied aspect of this challenge as the computational cognitive inference, and describe the major routes of computational approaches. To balance the potential optimism alongside the speed and scale of the data-driven era of cognitive science, we propose to inspect this trend in more empirical terms by identifying the operational challenges, societal impacts and ethical guidelines in conducting research and interpreting results from the computational inference in cognitive science.

\end{abstract}

% keywords can be removed

\section{Introduction}

There is a research trend in cognitive science that shifts from a top-down direction (guided by hypothesis-driven testing of cognitive theories) towards a bottom-up approach (enabled by data-drivcen pattern discovery of cognition-related properties). The emergence of high-throughput data collection techniques provides cognitive scientists rich research substances of labelled behavioral data, from one's digital traces on a social media, to large-scale crowdsourcing of experimental responses to well-defined cognitive tasks \cite{jones2017big}. Riding along the big data era of cognitive science is the advanced developments of artificial intelligence (AI) methods that is capable of performing components of cognitive functions at human-level or superhuman-level performances. With the new directions, comes new challenges. As the study of the essence, tasks and functions of cognition, how can we as cognitive scientists reshape the field using these new sources of data and new tools of analytical methods, such that it maintains a coherent core as the classical theory-driven studies of cognitive science?

\begin{figure}[tb]
% \vspace{-0.8cm}
\centering
    \includegraphics[width=\linewidth]{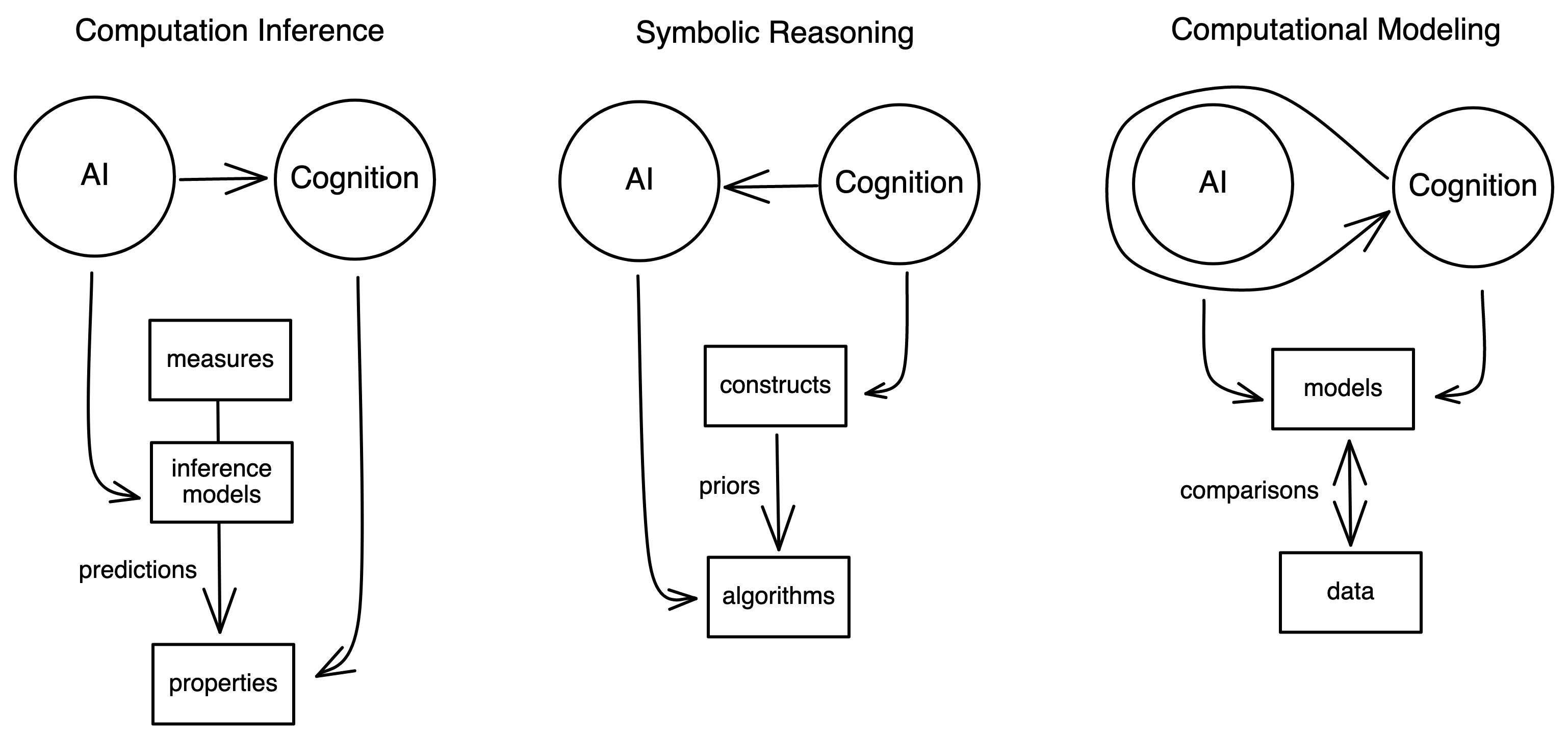}
\caption{The three interactions between the concepts of AI and those of the human cognition
}\label{fig:class}
\end{figure}

To better formulate this challenge, we categorizes the interactions between the concepts of AI and those of the human cognition into three main types (Figure \ref{fig:class}). First, we have the computational inference, the process of utilizing machine learning models as a prediction or inference engines to map from measurable signals to the cognitive properties. The second direction is to use the cognitive theory as a prior to build AI. This approach can be dated as early as the symbolic cognitive architectures in 1970s \cite{anderson2013architecture,simon1971human}, where major cognitive processes such as knowledge representation, memory, learning and control are explicitly mapped into computational components. A recent perspective piece by \cite{lake2017building} further points out the missing pieces in this direction: causality, intuition, compositionality and generalizability. The third direction, computational modeling, is to construct brain-computational models that mimic the certain properties of cognition or neurobiology and then compare them against experimental data. This approach is also termed as the cognitive computational neuroscience by \cite{kriegeskorte2018cognitive} which proposes to use task-performing computational models to test the cognitive processes against implementation-level hypotheses of neurobiologically plausible dynamic components, as well the main considerations in this interdisciplinary approach. 
Since the empirical considerations of the last two approaches has been discussed in the above work, we will focus our discussion to the first direction, the computational inference problem in cognitive science.

A simple distinction of the three approaches is that: 
the computational modeling of the mind refers to the use of computers to simulate the workings of the human mind; the symbolic reasoning of the mind is the ability of the mind to reason using symbols and abstractions; and the computational inference of the mind is the ability of the mind to make inferences based on computation.
One might relate this categorization to the Marr's distinction of the three levels of analysis: the computational theory, the algorithm, and the neurobiological or physical implementation  \cite{marr2010vision}. The algorithm level maps to building symbolic AI systems. The neurobiological implementation level maps to modeling computational mechanisms using biologically plausible model components. The objective of the computational inference, on the other hand, is to find the best surrogate models to predict the components and sub-processes of the computational theory. 

% A simple distinction of the three approaches is that: the computational modeling of the mind is focused on how the mind works; the symbolic reasoning of the mind is focused on how the mind can be used to solve problems; and the computational cognitive inference uses computational methods to infer the properties of the mind.

The top-down perspective of cognitive science decomposes complex cognitive processes into simplier subprocesses which has their own computational components. However, understanding these subprocesses (e.g. control, perception, learning and decision making) doesn't warrant a unified and coherent theory of cognition until very recently \cite{li2018gain}. Unlike the nascent development of finding a unified thoery in the basic research, the machine learning models which are independently tailored for specific inference tasks can already be readily used for applied research. As George Box the statistician wisely points out, ``all models are wrong, but some are useful'' \cite{box71}. Here we argue that, the computational cognitive inference is the cornerstone of applied cognitive science because they provide actionable inference anchors to cognitive concepts which can be efficiently binded with important downstream real-world applications and if taken responsible discretion, provide useful interpretable insights in clinical setting. 

In this work, we aim to address the conceptual challenges of inferring properties of cognition with computational models, in particular, from an applied point of view. First, we will formulate the research problem, the measurement approaches, and major routes of computational methods. What are the operational challenges of developing a data-driven cognitive science study? What are the societal impacts along with the analyses and interpretations of the experimental results? And what are the ethical guidelines to design, perform and analyze study of computational cognitive inference?  

\section{Problem Setting}

Here we first formulate the problem setting. From a statistical and signal processing point of view, the cognition is an unknown data generating process in the brain that ``generates'' the data, which we record externally with behavioral, imaging and physiological measurements. These measurements serve as an approximation to the underlying cognitive mechanism. The goal of computational inference is to find the most accurate computational account of the measured data to reflect the underlying cognitive properties. 

\subsection{Data Measurements}

If we consider certain cognitive property to reside in a high-dimensional manifold, the experimental measurements are the projections of this manifold to a low-dimensional space, corresponding to the cardinality of the measurement space. There are four primary sources of measurements.

% \begin{enumerate}[itemsep=10pt,topsep=10pt]
%     \item 
\textit{1. Behavioral data.} Many cognitive tasks measures task-specific behavioral data such as the task performance, reaction times, eye movements, and signatures of success and failures. The challenge of these behavioral measures is that, in order to characterize individual variation in their cognitive abilities and interpret their outcomes, the researchers need to design appropriate experiments with extensive considerations to rule out non-cognitive factors \cite{rowe2014measuring}.

% \item 
\textit{2. Biometric and brain activity data.} These behavioral data are often complemented by simultaneous recording  of brain activity and other biometric signals such as functional magnetic resonance imaging (fMRI), magnetoencephalography (MEG) and electroencephalography (EEG), which captures the spatio-temporal neural dynamics during the tasks. These physiological signals, if supported by easy-to-use wearable recording devices, can be used to build brain computer interfaces (BCI) \cite{abiri2019comprehensive,sitaram2007fmri} for various applications such as diagnosing neurological conditions \cite{alsaggaf2014directions} and facilitating rehabilitation of patients with neurodevelopmental disorders \cite{arpaia2020wearable}. 

% \item 
\textit{3. Self-reports.} The self-reports are another important measurement approach, extensively used in psychiatry, to connect a research instrument (which are rating scales designed to standardize the measurement and definition of psychiatric questions) to a cognitive condition \cite{thompson1989instruments}. However, it is observed that these self-reports of attitudes and behaviors can be strongly affected by the question wording, the rating scales, the questionnaire format, and the contexts of the research instrument \cite{schwarz1999self}. 
    
% \item
\textit{4. Linguistic and speech data.} Until recently, there is an growing interest in directly using the linguistic and speech signals as biomarkers to measure these cognitive properties, such that they can inform cognitive states, health conditions and therapeutic prediction in clinical setting. These features complement the self-reports and can enable real-world applications in large scales such as tracking population sentiment \cite{nasukawa2003sentiment}, providing topical context of psychotherapy \cite{lin2022neural}, diagnosing mental illness  \cite{cummins2015review}, monitoring of cognitive decline \cite{fraser2016linguistic}, and automatically recommending treatment strategies to therapists \cite{lin2022supervisor}.
% \end{enumerate}

\subsection{Computational Inference Models}

Along the notion of the underlying unknown manifold of the cognition properties, if the measurements are the projections to a low-dimension feature space, the computational inference methods provide reference frames within the feature space, in order to extract a accurate representation of the cognitive properties with reasonable scales and meaningful insights. Here we outline four classes of computational models for this inference problem.

% \begin{enumerate}[itemsep=10pt,topsep=10pt]
    % \item 
    \textit{1. Supervised learning.} One can utilize the available variables in the measurements to construct pairs of a measurement and a label. When analyzing behavioral data in cognitive tasks, the experimental conditions usually serve as a categorical label. Labels can also be other auxiliary variables such as age, gender, education level, country. In clinical studies, labels can be the doctor's diagnosis or evaluation of the severity of the psychiatric conditions. These evaluations can be a score obtained from clinically validated approaches, such as the self-report score of a patient in a clinical instrument inventory (e.g. a questionnaire with scales and keys for scoring). Machine learning methods have been readily available to extract out a mapping from the measurement to the label, by training upon a training set of measurement-label pairs. If the labels are categorical, a classification model can predict one or multiple classes as its output. If the labels are ordered or continuous, a regression model can predict a scalar number or vector corresponding to a validated quantity of cognitive properties.
    
    % \item 
    \textit{2. Unsupervised learning without priors.} In many cases, we don't have any labels for the quantity of the cognitive property of interest. In another case, we don't want to make assumptions which are constrained by the labels, and are simply interested in the underlying shape or structure of the measurements, in order to characterize the cognitive property into major classes or a ranked sequence. This is an intuitive assumption, because not only are computational models usually chosen based on how simply they can explain the empirical data in cognition (i.e. how little assumptions they make), the Occam's razor (or the rule of parsimony, or simplicity) also drive the majority of the cognitive processes    \cite{chater2003simplicity}.
    For instance, if we adopt the previous notion of data being a manifold, measurement being a projection, the principal component analysis (PCA) is a linear dimension-reduction technique that sets the reference frames such that the first few axes maximize the variance it can explain in the empirical data. After the dimension reduction, we can visualize and discover patterns in the data in a low-dimensional space (usually in a 2d picture). An example of this is to apply PCA to recognize facial expressions \cite{calder2001principal} by reframing the measurements onto a 2d continuous spectrum of emotions quantified by \cite{russell1980circumplex}.
    
    % \item
    \textit{3. Unsupervised learning with priors.} In clinical settings, especially in psychiatry, there are usually well-defined research instruments for cognitive properties, such as Beck depression inventory (BDI) \cite{jackson2016beck}, working alliance inventory (WAI) \cite{tracey1989factor}, and many others accepted as diagnostic tools in the Diagnostic and Statistical Manual of Mental Disorders (DSM) \cite{regier2013dsm}. These clinical instruments can serve as useful priors for the computational models. Like the computational methods, these instruments or inventories also offer a reference frame to position the measurements (which in this case is mostly self-reports) to a clinically validated spectrum of cognitive properties. As an example, in \cite{lin2022deep,lin2022deep2}, the authors compute a inference score using a deep learning-based sentence embedding to infer a score for the patient-therapist bonding directly using the working alliance inventory as a reference. Similarly, one can use multiple inventories together to find interactions among cognitive properties \cite{lin2022unsupervised}, where the effect of personality types on therapeutic relationship between the patients and therapists are characterized in a turn-level resolution. The benefit of this approach is that it doesn't require any training of the model on the measurement data, such that these inference models can be deployed to unseen data, as in this real-time annotation application of the above study \cite{lin2022voice}.
    
    % \item
    \textit{4. Generative learning.} In many cases, the measurements reflect a cognitive process that has an innate temporal dynamics, so the inference result would therefore be a sequence of outputs, which can be generated sequentially with a mechanistic or generative model. These generative models can learn about their environments from their past experiences and derive an optimal policy to recreate outputs given their current states. An example is the reinforcement learning (RL) model, which is reviewed as a comprehensive framework to study various critical cognitive processes, such as learning, representation and decision making \cite{botvinick2020deep}. Unlike the supervised learning approaches to mimic the cognitive processes inside human brains as in \cite{lin2022predicting,lin2022predicting2}, the reinforcement learning models to model cognitive processes are mechanistic and thus, offer generalization insights.
    To use reinforcement learning as a basis for inference, one can introduce neurobiological priors into the model architecture, such as the two-stream reward processing mechanism \cite{lin2019split} and the state dependency \cite{lin2020unified}. Since these priors are baked in learning parameters related to the labels (e.g. mental disorders modulated by reward processing), one can infer these labels by comparing the trajectories generated by these parameterized models with the ones in the measurements \cite{lin2020story,lin2021models}. One can also predict measurement trajectories with sequence-generating deep learning models, such as the recurrent neural networks and long short-term networks (LSTM) \cite{hochreiter1997long}. Analyzing the weight distribution of these approximate inference engines provides an interpretable insight on the underlying cognitive process (e.g. the distribution of decision making strategies adopted by individuals with different psychological propensities) \cite{lin2022predicting}. Lastly, one can represent the probability of each configuration of the cognitive concepts with models using variational inference \cite{galdo2020variational}, such as the generative adversarial networks (GAN) \cite{goodfellow2020generative} or variational auto-encoder (VAE) \cite{kingma2013auto}.
% \end{enumerate}

\section{Operational Challenges}

There are many operational challenges of building computational inference tools for cognitive science, such as the lack of a clear definition of what constitutes the cognitive properties, the difficulty of accurately modeling the complex interactions between underlying cognitive processes and their elicited behaviors, and the requirements of creating models that are both predictive and reproducible. Given the limited amount of data available for training computational models, there is usually a risk of overfitting when using the trained computational models to infer cognitive properties, which can lead to inaccurate results. Additionally, computational inference methods may be sensitive to small changes in the data, making it difficult to replicate results. Some empirical qualities that we should aim for when developing and deploying computational cognitive inference models include its validity, reliability, and generalizability. We propose five practical considerations to tackle these operational challenges.

% \begin{enumerate}[itemsep=10pt,topsep=10pt]
    % \item 
    \textit{1. Evaluate the technological readiness and real-world generalization.} The limitations of many studies in computational cognitive inference reside in the difficulty of a systematic clinical validation and a uncertain future expectation of the technological readiness for patient care and therapeutic decision making approved by authorities. For instance, it was recently shown that despite the high predictability of statistical learning-based methods in analyzing large datasets in support of clinical decisions in psychiatry, existing machine learning solutions is highly susceptible to overfitting in realistic tasks which has usually a small sample sizes in the data, missing data points for some subjects, and highly correlated variables \cite{iniesta2016machine}. 
    These properties in real-world applications can limit the out-of-sample generalizability of the inference results. 
    
    \textit{2. Incorporate model validation into the experimental design.} How can we interpret and validate the results from computational models of cognitive properties if there is no golden standards (e.g. from the clinical standpoint) to compare with? One approach is to use computational models to generate predictions about how people with various cognitive impairments might perform on various tasks. These predictions can then be compared to empirical data to see how well the model accounts for the real-world performance. Another approach is to use computational models to generate hypotheses about the mechanisms underlying cognition, which can then be tested with experiments which collect empirical data to further compare with the prediction.

    % \item
    \textit{3. Consider the tradeoff between the predictive power and model complexity.} Other than using the inference results to generate new experimental hypotheses, another effective way to increase the generalizability of computational models is to design simpler computational models in the first place. Building simpler models can help researchers to identify the key variables at play in a particular observed phenomenon, and to understand the relationships between those variables. Additionally, simpler models are often easier to communicate to other researchers and to the general public, such that oversights can be caught in early phase of system deployment of these cognitive inference engines. 
    
    % \item
    \textit{4. Be mindful of the publication bias.} Publication bias often occur when the results of a study are more likely to be published if they report a positive or significant finding. This can lead to a distorted view of the predictive power of a cognitive properties, or the effectiveness of a treatment or intervention based on such inference results. It has led to a number of instances where machine learning (ML) models were deployed in the healthcare industry given a positive results reported in publications which evaluated on smaller datasets, but later find these performance predictions are overly optimistic \cite{wong2021external}.
    To avoid publication bias, the researchers can perform a systematic review to identify, select, and critically appraise all of the evidence relevant to the cognitive processes that the computational solutions aim to investigate. Meta-analysis is another useful technique that combines the results of multiple studies to estimate the effect of a treatment or exposure (which in our case, are the performance estimates of different predictive models on the same cognitive process). A community effort could be setting up a clinical trial data repository or registry, such that information about ongoing and completed clinical trials, as well as information about published and unpublished studies, can be shared within the research community. Then during the deployment phase in the real-world application, one can use a search filter to select a set of criteria in this shared database to identify studies and results that meet certain criteria, such as date, study design, cognitive processes and population, in order to make more reasonable estimates of the inference accuracy.
    
    % \item
    \textit{5. Create interpretable computational models.} Making the computational models interpretable have three benefits. First, other than its predictive power for applied research, these models can provide auxilary information about how minds works. If we can see how the models work, we can better understand the cognitive processes that they are trying to simulate. Additionally, interpretable models can be useful for debugging and improving existing models. If we can see what is going wrong with a model, we can more easily fix the problem. Finally, interpretable models can be used to create new models. By understanding how existing models work, we can build new models that are more accurate and realistic to predict certain cognitive elements. To build an interpretable model, we can first design the models to be as simple as possible while still capturing the relevant phenomena. In the model documentations, we should use clear and intuitive terminology when describing the models, provide detailed explanations of how the model works, and define the goals of the model and the phenomena it is meant to explain. In addition, provide methods to visualize the model in a way that is easy to understand. From the software engineering standpoint, one should include unit tests, extensive simulations and analyses to ensure that the model behaves as intended. Finally, we should compare the cognitive inference systems to see if it can be explained in light of existing psychological theories and empirical data.
% \end{enumerate}

% A recent work \cite{Wang2022}

\section{Societal Impacts}

% Cognitive science research often has two impacts to modern society. The first one is a philosophical one. 

The society that we know of is changing at a rapid speed. Investigating the results from the computational inference models of cognitive constructs can augment our understanding of human cognition and behavior. The predictive power of one's cognitive properties can improve decision-making in various fields such as developing more personalized treatment plans in clinical medicine, recommending more suitable services for individual users in business, and creating more reasonable laws taken into account the societal propensity of a certain behavior. These tools can also create more efficient and effective educational practices. Although many of these inference engines are powered by artificial intelligence, the knowledge we learned from cognitive science can in turn help design better artificial intelligence systems for other application domains.

While the usage of data-driven computational approaches into cognitive science can have huge positive impacts to how we view ourselves and the world around us, and sometimes, these new understanding can be oversimplified and even misleading. Some societal concerns of misinterpreting results from computational cognitive inference models include issues of privacy, security, and fairness. For example, if a cognitive inference model is used to make inferences about an individual's mental state or abilities, there could be implications for that individual's privacy or civil rights. Additionally, if a cognitive science model is used to make inferences about a group of people, there could be implications for the security of the group (if the model is used to make predictions about future behavior, for example) or for the fairness of the group (if the model is used to make decisions about who to hire, for example). We propose four suggestions to consider to guide our understanding of the findings and applications from computational cognitive inference.

% \begin{enumerate}[itemsep=10pt,topsep=10pt]
% \item 
\textit{1. Consider the proper role of computational inference engines.} Digital applications from computational cognitive inference should choose a right place in societal setting. Take mental health as an example. Till now, there are still a severe global shortage of workforce in mental health \cite{olfson2016building,satiani2018projected}.
This demand gap has recently increased due in part to the impact that the COVID-19 pandemic has had on everyone's mental health \cite{wang2020investigating}. Due to the years of continuous learning and supervised training that each licensed therapist needs, current educational systems and training programs cannot keep up with this trend. A proper computational inference framework should not aim to replace existing work force of psychiatrists or therapists, but to assist them. In education setting, having an interpretable models that can inform the next-generation psychiatrists about the strategies adopted by experienced therapists. In this example, the usage of automatic computational tools can potentially alleviate this societal issue in both assisting and educating junior psychiatrists in a more scalable way, instead of taking an adversarial role of replacing these important occupations.
% Similarly, we should consider 

% \item 
\textit{2. Define and redefine cognitive constructs in an iterative process.} Cognitive constructs are mental representations of external reality, which are created by the interaction between our senses and our brains. Because our brains are always changing and growing according to the changing world around us, our definitions of the cognitive constructs are also always changing and growing. As new computational models are proposed to infer cognitive elements from different perspectives, new constructs can emerge from the patterns discovered by computational models. Computational inference of cognitive properties can help us redefine cognition by providing a more detailed, comprehensive and accurate understanding of how the mind works. This could lead to a better definition of cognition that takes into account the various interacting processes involved in mental activity. For instance, by understanding the algorithms and processes that underlie cognition through interpreting our inference models, we can design better mechanistic hypothesis of how the mind works and identify new cognitive properties that were previously unidentified or poorly understood. This could help to expand and improve our understanding of cognition. Additionally, computational inference can help us develop new ways of measuring and assessing cognition, which can lead to a more refined understanding of the mind.

% \item 
\textit{3. Determine the right questions to answer.} The computational inference aims to depict the data generating mechanisms of how data (the neural space) is processed in the background (the cognition space) to produce behavior (the measurement space). As a result, the knowledge informed by model is constrained by the measurement data it have access to, which is further constrained by the experimental design and technique. Therefore, it is important to ask the right question in the experimental phase that matches the specific goal of the computational model and the cognitive phenomenon being studied. To find the right question, one should identify the key variables and processes that are relevant to the phenomenon of interest, formulate testable hypotheses about how these variables and processes interact to produce the observed behavior, choose the model structure and parameters that will allow us to test those hypotheses, confirm that the model design is consistent with what is known about the cognitive phenomenon from experimental and observational data, and finally test the trained model against new data to see if it can accurately predict behavior.

% \item 
\textit{4. Acknowledge the multi-dimensional complexity of cognition.} Cognition is a complex and multi-dimensional phenomenon. Existing datasets on which computational inference are trained are usually from controlled experiments that studied one primary cognitive process. This discrepancy can limit the generalization power of the cognitive inference findings in the field setting, where multiple sensory and cognitive systems are at play. One way to approach this question is to consider the various ways in which cognition can be studied and directly bake them into the model design. For example, cognition can be studied from a psychological perspective, looking at how mental processes such as memory and attention operate. Alternatively, cognition can be studied from a neurological perspective, looking at how the brain supports these mental processes. Finally, as the main focus of our study, when cognition is studied from a computational perspective, one may incorporate the neurobiological and psychological priors into the model architecture or parameters, and then look at how these biologically plausible computational models can simulate or explain the cognitive phenomena. Each of these perspectives can provide valuable insights into the nature of cognition, and by considering all of them together we can begin to build up a more complete picture of this complex phenomenon. Without this complete picture, we should acknowledge the limitations of existing cognitive inference results with respect to their implications to population behaviors and social phenomena.
% \end{enumerate}

% 1. Increasing our ability to process and analyze information: AI can help us process and analyze information more effectively by providing us with new ways to organize and visualize data.

% 2. Enhancing our memory and learning: AI can help us improve our memory and learning by providing us with new ways to store and retrieve information.

% 3. Helping us make better decisions: AI can help us make better decisions by providing us with new ways to evaluate data and identify patterns.

% 4. Allowing us to interact with machines in new ways: AI can help us interact with machines in new ways that are more natural and efficient.

% 5. Enabling us to create new things: AI can help us create new things that would not be possible without its assistance.

\section{Ethical Guidelines}

Some ethical implications of building computational inference models for cognitive science include the potential for AI bias and discrimination, the need for transparent and explainable AI, and the potential for misuse of AI. These implications can exist in the full life cycle of the computational inference (data collection, model development, system training and deployment). One concern is that these models could be used to make decisions about people without their knowledge or consent. Another concern is that the models could be used to manipulate people's behavior. These methods, even if used in acceptable contexts, could be used to unfairly stereotype or discriminate against certain groups of people. As more and more successful applications of artificial intelligence are deployed in clinical domains related to treating and supporting the cognitive aspects of human minds, there are many ethical considerations the practitioners of machine learning should acknowledge and take into considerations. 

% \begin{enumerate}[itemsep=10pt,topsep=10pt]
    % \item 
    \textit{1. Respect the privacy and anonymity of human subjects.} When working with the patient data, security and privacy are the top priority in applied setting. Following the suggestion of best practices from \cite{matthews2017stories}, all data from human subjects should be properly anonymized with pre- and post-processing techniques before the analyzing. For instance, if readers can back trace the identity of the subjects given a case study, they are exposed to marginalization and stereotypes even if there is no causalities studied between the cognitive properties and identity-related variables (such as a specific gender or ethnicity). 
    
    % \item 
    \textit{2. Be mindful of the risk of introducing unexpected bias.} Another ethical boundary to maintain is to make sure that the AI systems we use to diagnose, interpret and predict the cognitive properties don't lead to increased risks to the patients. This requires both the practioners and machine learning researcher to fully aware of potentially bias and ethical challenges, such as gender bias, language-related ambiguity and ethnicity-related mental illness connections \cite{chen2019can}, in order to deploy the AI system in a responsible and safe way. One solution is to pay attention to how data is collected and labeled, and make sure that data is coming from a variety of sources that are representative of the population we are trying to model. We can inspect the data after they are collected for any patterns that could be indicative of bias, and if we find any, we can attempt to determine why they exist and whether they are likely to impact your results. From the modeling aspects, we can use a variety of AI techniques to train the computational models, which can reduce the chances that any one specific technique is introducing the bias. We can also evaluate the inference results against a variety of different criteria with consultations to different domain experts, which can help ensure that the inference models are not introducing bias in any one particular area.

    % \item 
    \textit{3. Accept the limitations that many computational models are proof of concepts.} Emerging techniques in wearable devices, digital health records, brain imaging measurements, smartphone applications and social media are gradually transforming the landscape of the monitoring and treatment of cognition-related mental illness. However, most of these attempts are merely proof of concepts, and requires substantial caution to prevent from the trap of over-interpreting preliminary results \cite{graham2019artificial}. These proof of concepts inform us whether using a specific computational approaches to infer a certain cognitive construct is feasible or not, while in applied research, we aim to create validated engineering systems that have been proven to work effectively across a variety of plausible conditions and are ready to be used in a real-world setting. One danger of mistaking a proof-of-concept digital health tool (e.g. for mental health) as a technologically ready system is that it may not be able to accurately diagnose or treat a person's condition, and provide the person with the necessary information to make informed decisions about their health, which as we have already pointed out the red line in the second point, can introduce underestimated risks for the person (e.g. missing a time-sensitive treatment due to a false estimate of disease progression). 
    
    % \item 
    \textit{4. Bring the public into discussion at early stage.} As we introduce new AI systems into cognitive modeling space, one way of promoting this awareness is to engage the public in discussions about the proper usage of artificial intelligence in cognitive science, mental health and well beings. One of the main message to communicate is that we should avoid optimistically expecting the AI to be the ``domain expert'' \cite{carr2020ai}. The discussion of what is right and what is not should be at every step of the development of the AI system, instead of only at the end of the project cycle as an after thought. From the applied side, being transparent about the computational inference models we use and the results we get can also help build trust with the audience and allow others to audit the inference results for bias.
    % , such as ones pointed out in the second point.

\section{Conclusions and future outlooks}

The computational cognitive inference is the process of making inferences about mental states and cognitive processes based on computational models. There is potential in this approach to use data-driven techniques  to interpret cognitive and psychological constructs, but it is not clear how successful this will be in real world setting. There is a risk that these computational models will not be able to accurately interpret these constructs, which could lead to inaccurate conclusions. Still, the future of using emerging techniques such as AI to interpret cognitive elements is very promising. With the help of AI, psychologists and psychiatrists will be able to more accurately diagnose and treat mental disorders. Additionally, AI can be used to help people with cognitive disabilities to better understand and communicate their thoughts and emotions. Another area related to fundamental science is to use these inference models to further understand how the brain represents and processes information. This could lead to a better understanding of how the brain works in normal and abnormal states, and could potentially be used to develop new treatments for mental disorders. Operationally, it is important for cognitive scientists and machine learning practitioners to consider the technological readiness of the inference methods, the experimental design to allow for iterative evaluation, and the complexity and interpretability of the inference model. When cognitive inference is validated and deployed in applied systems, we need to consider its societal implications by position it in the right role and acknowledge the cognitive complexity of real-world scenarios. Finally, we should follow a responsible ethical guideline to be transparent about the approach, realistic about the expectations, and mindful of the potential bias.

\bibliographystyle{unsrt}
\bibliography{main} 

\end{document}